# Different reconstruction pathways toward superconductivity in $TaRhTe_4$ and $TaIrTe_4$ Weyl semimetals

Jinyu Zhao[1*], Yang Fu[1*], Congcong Le[2*], Shuaihang Sun[1], Shu Cai[1], Yang Ding[1], Qi Wu[3], Hechang Lei[4]†, Jiangping Hu[3]†, Lili Zhang[5], Cedomir Petrovic[1,6], Liling Sun[1, 3]†

[1]*Center for High Pressure Science & Technology Advanced Research, Beijing 100193, China*
[2]*Hefei National Laboratory, Hefei 230088, China*
[3]*Institute of Physics, Chinese Academy of Sciences, Beijing 100190, China*
[4]*Department of Physics, Beijing Key Laboratory of Optoelectronic Functional Materials and Micro-Nano Devices, Renmin University of China, Beijing 100872, China*
[5]*Shanghai Synchrotron Radiation Facility, Shanghai Advanced Research Institute, Chinese Academy of Sciences, Shanghai 201204, China*
[6]*Shanghai Key Laboratory of Material Frontiers Research in Extreme Environments, Shanghai Advanced Research in Physical Sciences, Shanghai 201203, China*

Pressure can drive Weyl semimetals toward superconductivity through qualitatively distinct reconstructions of their lattices and normal-state electronic structures. Here, we report the first observation of superconductivity in compressed $TaRhTe_4$. This finding enables a direct comparison of the distinct reconstruction pathways leading to superconductivity in $TaRhTe_4$ and the previously studied $TaIrTe_4$, both of which belong to the $TaXTe_4$ (X = Rh, Ir) family of type-II Weyl semimetals. For $TaRhTe_4$, high-pressure electrical-resistance, Hall effect, and magnetoresistance measurements, together with synchrotron X-ray diffraction and first-principles calculations, reveal a superconducting transition that emerges near 20 GPa, with onset $T_c$ increasing to approximately 2.6 K at 65.2 GPa and zero resistance achieved above 63 GPa. The onset of superconductivity coincides with a progressive lattice distortion, a strong suppression of the positive magnetoresistance, and a continuous decrease of the Hall coefficient toward zero. Calculations further show that additional electron-like bands cross the Fermi level ($E_F$) and that $N(E_F)$ increases upon compression. This evolution contrasts with $TaIrTe_4$, where the Hall coefficient initially increases before reversing its pressure dependence near the superconducting threshold, while the structural anomaly is confined to a narrower pressure interval. This comparison indicates that superconductivity in the $TaXTe_4$ family is not tied to a unique critical pressure or a single Fermi-surface reconstruction, but can emerge through distinct material-specific pathways once pressure sufficiently reconstructs the low-carrier Weyl-semimetal-derived state into a multiband metallic regime.

Weyl semimetals host topologically nontrivial band crossings and exhibit unusual transport properties associated with their low-energy electronic structure [1–5]. $TaRhTe_4$ is a layered van der Waals material crystallizing in the orthorhombic space group $Pmn2_1$ and has been identified theoretically as a bulk type-II Weyl semimetal [6–8]. Its low-energy states are dominated by Ta-($d$) and Te-($p$) orbitals, and the electronic topology is sensitive to structural parameters such as the interlayer configuration [7,9–11]. Recent magnetotransport measurements have revealed negative longitudinal magnetoresistance consistent with a chiral-anomaly response [5,9], making $TaRhTe_4$ a useful platform for probing the evolution of a Weyl-semimetal state under external tuning.

Pressure provides a clean and continuous means of tuning lattice structures and electronic states without introducing chemical disorder. Compression can tune interatomic distances, orbital hybridization, carrier balance, and Fermi-surface topology, and can induce superconductivity in a variety of Weyl and topological semimetals. In $WTe_2$ and $MoTe_2$, pressure-induced superconductivity is accompanied by pronounced changes in normal-state transport and electronic structure [12–15], whereas in TaP superconductivity appears together with a pressure-induced structural transition [16,17]. Of particular relevance here is $TaIrTe_4$, a close structural analogue of $TaRhTe_4$. In $TaIrTe_4$, superconductivity emerges near 23.8 GPa as the positive magnetoresistance collapses and the lattice becomes distorted; at the same time, the pressure dependence of the positive Hall coefficient reverses above the superconducting threshold [18]. Related pressure-induced superconductivity has also been reported in $NbIrTe_4$ [19–21]. These results raise a broader question: do members of the $TaXTe_4$ family reach superconductivity through a common electronic-reconstruction pathway, or can distinct reconstruction pathways lead to a similar superconducting state? The high-pressure evolution of $TaRhTe_4$ provides a direct test of this question. Here, we combine high-pressure electrical-resistance, Hall effect, and magnetoresistance measurements with synchrotron X-ray diffraction and first-principles calculations to establish the pressure evolution of $TaRhTe_4$ and compare its reconstruction pathway with that of $TaIrTe_4$.

Before the high-pressure experiments, we performed ambient-pressure X-ray diffraction measurements on $TaRhTe_4$ samples obtained from different batches. These measurements confirm that both samples crystallize in the orthorhombic (OR) structure with space group $Pmn2_1$ (see Supplementary Information (SI) [22]). We then conducted high-pressure resistance measurements on these two single crystals. Figures 1(a) and 1(d) show the temperature dependence of the electrical resistance normalized by its value at 300 K, $R/R_{300K}$, for the two single-crystal samples, S#1 and S#2, under pressures up to 44.3 and 65.2 GPa, respectively. Throughout the investigated pressure range, the normal-state resistance decreases upon cooling, indicating that $TaRhTe_4$ retains overall metallic transport behavior under compression. However, the normalized resistance $R/R_{300K}$ exhibits a pronounced nonmonotonic pressure dependence. For S#1, the low-temperature $R/R_{300K}$ increases progressively with pressure up to 44.3 GPa, indicating a gradual weakening of the temperature dependence of the normal-state resistance. A similar trend is observed in S#2, where the low-temperature $R/R_{300K}$ increases with pressure up to 45.6 GPa. Upon further compression, this trend reverses, and the low-temperature $R/R_{300K}$ decreases, indicating that the temperature dependence of the metallic resistance strengthens again. This nonmonotonic evolution reveals substantial changes in the normal-state transport properties under pressure. Notably, this reversal approximately coincides with the completion of the OR to distorted-OR (d-OR) structural transition discussed below.

A closer examination of the low-temperature data reveals pressure-induced superconducting transitions in both samples. As shown in Fig. 1(b), no obvious transition is observed in S#1 at 15.6 GPa, whereas a weak resistance downturn develops at 19.7 GPa near the lowest accessible temperature. Upon further compression, the transition becomes progressively more pronounced and shifts to higher temperatures. Although the resistance approaches zero upon further compression, a finite residual resistance remains in S#1 down to the lowest measured temperature. In S#2, a similar resistance drop emerges and becomes increasingly pronounced with pressure [Fig. 1(e)].

A complete zero-resistance state is achieved at 63.3 GPa and persists to 65.2 GPa. Using the normal-state extrapolation criterion, the superconducting transition temperature ($T_c^{onset}$) reaches 2.3 K at 44.3 GPa for S#1 and 2.6 K at 65.2 GPa for S#2. These results demonstrate the progressive development of pressure-induced superconductivity in $TaRhTe_4$, culminating in a robust zero-resistance state at the highest pressures investigated.

The magnetic-field dependence of the two samples provides further evidence of their superconductivity. As illustrated in Figs. 1(c) and 1(f), the superconducting transitions shift systematically to lower temperatures and becomes progressively suppressed as the magnetic field increases, as measured at 44.3 GPa for S#1 and 65.2 GPa for S#2. This magnetic-field response confirms that the pressure-induced resistance drops originate from superconductivity. Fitting the temperature-dependent upper critical field to the Ginzburg-Landau (GL) model yields zero-temperature values $\mu_0H_{c2}(0)$ of approximately 2.1 T at 44.3 GPa for S#1 and 2.8 T at 65.2 GPa for S#2. For *H//c*, the in-plane GL coherence lengths, derived using ξab(0) = [Φ0/(2πμ0Hc2(0))]1/2, are about 12.5 nm and 10.8 nm, respectively. These values are comparable to the approximately 12 nm GL coherence length reported for the pressure-induced superconducting state of $NbIrTe_4$ [20].

To investigate the pressure evolution of normal-state electronic transport, Hall resistance and magnetoresistance measurements were performed on S#2 at 10 K. As shown in Fig. 2(a), $R_{xy}$ is approximately linear in magnetic field over the measured range, consistent with previous ambient-pressure measurements [9,23]. The positive Hall slope is relatively large at low pressures but decreases markedly upon compression and fluctuates around zero in the high-pressure regime. This monotonic suppression differs from the behavior reported for $TaIrTe_4$, where the positive Hall coefficient ($R_H$) initially increases and reverses its pressure dependence only near the superconducting threshold [18]. Because $TaRhTe_4$ is a multiband semimetal, the reduction of the effective Hall response cannot be reliably interpreted in terms of a single-band carrier density. Instead, it signals a substantial redistribution of electron- and hole-like transport contributions, potentially involving changes in both carrier balance and mobility. The distinct pressure dependences of $R_H$ in $TaRhTe_4$ and $TaIrTe_4$ therefore indicate that the two compounds undergo distinct normal-state multiband reconstructions before entering their superconducting states.

The transverse magnetoresistance (MR) exhibits similarly pronounced pressure dependence. As shown in Fig. 2(b), $TaRhTe_4$ exhibits positive MR throughout the pressure range investigated. At 4.1 GPa, the MR reaches approximately 4.6% at 7 T and is progressively suppressed with increasing pressure, becoming smaller than approximately 0.46% above 20 GPa. The strongest reduction therefore occurs primarily below and near the pressure at which the low-temperature superconducting transition first appears. A similar collapse of positive magnetoresistance (MR) accompanies the onset of superconductivity in $TaIrTe_4$, though its low-pressure MR is considerably larger and drops to nearly zero at around 25 GPa [18]. Thus, the suppression of positive MR is a common feature of the two compounds, whereas the accompanying Hall evolution is not. This combination suggests that pressure strongly reconstructs the compensated or multiband transport state in both systems, with distinct trajectories in carrier balance and mobility.

To quantify the evolution of the field dependence, the MR curves were empirically fitted using MR=A$B^{\alpha}$ over the field range from 0 T to 8 T (see SI [22]). The perfector $A$, which characterizes the overall MR magnitude, decreases from approximately 0.15 at 4.1 GPa to 0.01 at 45.6 GPa. Meanwhile, $\alpha$ remains between 1.74 and 1.80 below 20 GPa and then gradually decreases from about 1.67 to 1.39 between 20 and 45.6 GPa. This evolution indicates that pressure not only suppresses the magnitude of the MR but also modifies its magnetic-field dependence from a nearly quadratic response toward a more subquadratic form. However, because the MR signal becomes very small at higher pressures, the variation of the exponent $\alpha$ cannot be assigned to a specific microscopic mechanism. Overall, the simultaneous suppression of the Hall response and MR demonstrates a pronounced modification of normal-state multiband transport before and through the emergence of superconductivity.

To examine whether the pronounced evolution of electronic transport is accompanied by structural changes, synchrotron X-ray diffraction measurements were performed up to 66.0 GPa at beamline 15U of the Shanghai Synchrotron Radiation Facility. As shown in Fig. 3(a), the diffraction patterns evolve systematically upon compression, with most reflections shifting toward higher angles, consistent with lattice contraction. In the intermediate-pressure range, however, the peak positions show subtle deviations from simple monotonic behavior, as further illustrated by the pressure dependence of the selected *d*-spacings in Fig. 3(c).

Below 20 GPa, all diffraction peaks can be well indexed by the ambient-pressure OR phase with space group $Pmn2_1$. The corresponding lattice parameters *a*, *b*, and *c*, together with the unit-cell volume *V*, decrease smoothly with increasing pressure [Fig. 3(b)], with no evidence of the first-order structural transition in this pressure range.

Above 20 GPa, however, several reflections can no longer be satisfactorily described by the OR structure. The pressure evolution of the selected *d*-spacings identifies two characteristic pressure regions near 20 and 40 GPa [Fig. 3(c)], indicating a progressive distortion of the OR lattice. In particular, the pressure-dependent changes in several reflections within the intermediate-pressure region suggest that the structural evolution is not merely a simple uniform lattice compression. We therefore describe the intermediate-pressure range as an OR-to-d-OR crossover regime, denoted OR + d-OR. Above 40 GPa, the diffraction patterns are dominated by the distorted structure (d-OR). Since the available diffraction data do not allow an unambiguous determination of the detailed high-pressure crystal structure, d-OR here should not be taken as evidence for a distinct crystallographic phase. Importantly, this approximately 20-40 GPa crossover is substantially broader than the lattice anomaly reported for $TaIrTe_4$, which is concentrated near its superconducting threshold around 23-27 GPa [18]. The two compounds thus differ not only in how their Hall response evolves but also in the pressure range over which their lattice reconstruction occurs.

The pressure dependence of electrical resistance provides an additional signature of this structural evolution. As shown in Fig. 3(d), the resistance at 5 K exhibits a weak nonmonotonic pressure dependence in both S#1 and S#2 and reaches a broad maximum near $P_{c1}$ before decreasing markedly upon further compression. Although the resistance at 300 K decreases continuously with pressure, its pressure dependence also changes noticeably across the same pressure region. Near $P_{c2}$, the resistance at both temperatures shows a substantially weaker pressure dependence, particularly in S#2. These reproducible features broadly coincide with the structural crossover regions identified by X-ray diffraction, indicating that the normal-state transport evolves concurrently with the pressure-induced lattice reconstruction.

We summarize our high-pressure results for $TaRhTe_4$ in Fig. 4. Below $P_{c1} \sim 20$ GPa, the sample remains in the ambient-pressure OR structure, and no superconducting transition is detected above 1.5 K. Near $P_{c1}$, the d-OR structure begins to develop, accompanied by the emergence of superconductivity. Within the OR + d-OR crossover regime (20-40 GPa), the value of $T_c^{onset}$ increases gradually with pressure. Above $P_{c2}$ (~ 40 GPa), the structural evolution toward the d-OR phase is essentially complete, while superconductivity persists and $T_c^{onset}$ continues to increase gradually, with a zero-resistance state first observed at 63.3 GPa and sustained up to 65.2 GPa. The smooth evolution of $T_c^{onset}$ across the two critical pressures indicates that superconductivity develops progressively through the pressure-induced lattice distortion rather than arising from an abrupt phase transition.

The normal-state Hall and MR responses exhibit pronounced changes over the same pressure range. As shown in Fig. 4(b), the effective Hall coefficient $R_H$ (defined in the SI [22]) decreases rapidly from approximately 0.059 $cm^3/C$ at 4.1 GPa and approaches zero near the high-pressure side of the OR + d-OR crossover regime. It subsequently fluctuates around zero in the d-OR phase. Meanwhile, the MR decreases dramatically from 4.1 GPa to 20 GPa and remains strongly suppressed at higher pressures [Fig. 4(c)]. Thus, the principal reductions in $R_H$ and MR occur before and through the onset of superconductivity and the structural crossover.

The combined transport and structural results show that superconductivity in $TaRhTe_4$ develops as part of a broad pressure-driven reconstruction of the normal state.

Below $P_{c1} \sim 20$ GPa, the system retains the ambient-pressure OR structure, and no clear superconducting transition is detected above 1.5 K. However, superconductivity below the experimental temperature limit cannot be excluded. As pressure approaches $P_{c1}$, both the Hall response and magnetoresistance are strongly suppressed, while a lattice distortion begins to develop and the first superconducting resistance drop appears. Superconductivity then evolves continuously through the OR + d-OR crossover regime and into the predominantly d-OR state above $P_{c2} \sim 40$ GPa, with $T_c^{onset}$ gradually approaching 2.5–2.6 K. The concurrent suppression of $R_H$ and MR thus indicates a pronounced reconstruction of normal-state multiband transport.

The DFT calculations provide a qualitative electronic picture of this evolution as pressure approaches $P_{c1}$. The calculated band structures and orbital-resolved density of states (DOS) at 0, 6.7, 12.4, and 20 GPa are shown in Figs. S4 and S5 of the SI [22], while the pressure dependence of the total DOS at the Fermi level is displayed in the inset of Fig. 4(c). At ambient pressure, the states near $E_F$ are dominated by Te-*p* and Ta-*d* orbitals, and the calculated DOS exhibits a pronounced minimum around $E_F$, consistent with a low-carrier-density semimetallic state. With increasing pressure toward 20 GPa, the low-energy bands evolve substantially, additional electron-like states cross the Fermi level, and $N(E_F)$ increases. These changes are consistent with the development of a more complex multiband Fermi surface and with the observed Hall evolution. Together with the suppressed MR, they indicate a substantial reconstruction of the normal state. Notably, the most pronounced calculated electronic evolution occurs within the same pressure range over which the lattice distortion first becomes apparent experimentally, suggesting a close coupling between electronic and structural degrees of freedom. Within the present calculations, which extend to 20 GPa, the increase in $N(E_F)$ may increasingly favor superconductivity; however, this correspondence should be regarded as qualitative and does not by itself establish the microscopic pairing mechanism.

A comparison between $TaRhTe_4$ and $TaIrTe_4$ reveals a common endpoint reached through different routes. Superconductivity in both compounds emerges only after pressure has strongly modified the low-pressure Weyl-semimetal-derived normal state. In both compounds, the onset of superconductivity is accompanied by lattice distortion and a pronounced suppression of positive magnetoresistance. The electronic reconstruction, however, differs markedly. In $TaIrTe_4$, the positive Hall coefficient first increases and then reverses its pressure dependence near the superconducting threshold [18], whereas in $TaRhTe_4$ it decreases continuously toward zero. The structural evolution is also broader in $TaRhTe_4$, extending over approximately 20-40 GPa, rather than being concentrated near the onset pressure. These contrasts argue against a universal critical pressure or a unique Fermi-surface reconstruction as a prerequisite for superconductivity in the $TaXTe_4$ family. Instead, they suggest that superconductivity becomes favorable once pressure drives the low-carrier Weyl-semimetal-derived state into a sufficiently reconstructed multiband metallic regime, with the detailed trajectory governed by material-specific band alignment, orbital hybridization, carrier mobilities, and lattice response. For $TaRhTe_4$, the DFT calculations provide a microscopic picture of this trajectory: compression toward 20 GPa introduces additional electron-like states at $E_F$ and increases $N(E_F)$, consistent with the observed Hall suppression and the development of a more complex multiband Fermi surface. Whether the Weyl topology survives in the high-pressure distorted regime remains an open question for the broader family and for $TaRhTe_4$ in particular.

In summary, the high-pressure evolution of $TaRhTe_4$, first observed in this study, reveals a reconstruction pathway toward superconductivity that is distinct from that previously established for $TaIrTe_4$. In $TaRhTe_4$, the superconducting transition emerges near 20 GPa as the Hall response and positive magnetoresistance are strongly suppressed and the lattice begins a broad OR-to-d-OR reconstruction extending to approximately 40 GPa. Superconductivity then persists in the predominantly distorted regime, with $T_c^{onset}$ reaching approximately 2.6 K at 65.2 GPa. First-principles calculations show that the compression toward 20 GPa reconstructs the low-energy electronic structure as additional electron-like states cross the Fermi level and $N(E_F)$ increases. By comparison with the high-pressure behavior of $TaIrTe_4$, we find that superconductivity in $TaRhTe_4$ and $TaIrTe_4$ is not tied to a unique Fermi-surface evolution or universal critical pressure, but instead to the formation of a sufficiently reconstructed multiband metallic state. Although the microscopic pairing mechanism and the fate of the Weyl topology at high pressure remain unresolved, this comparative picture provides a broader framework for understanding pressure-induced superconductivity in the $TaXTe_4$ Weyl-semimetal family.

*These authors contributed equally to this work.
Correspondence and requests for materials should be addressed to: Hechang Lei (hlei@ruc.edu.cn), Jiangping Hu (jphu@iphy.ac.cn) and Liling Sun (liling.sun@hpstar.ac.cn or llsun@iphy.ac.cn).

**Acknowledgements**

This work was supported by the National Key Research and Development Program of China (Grants Nos. 2022YFA1403900, 2021YFA1401800, 2022YFA1403800, 2023YFA1406500) and the National Natural Science Foundation of China (Grant Nos. 12625409 and 12274459). J.P.H. is supported by the Ministry of Science and Technology (Grant No. 2022YFA1403901), National Natural Science Foundation of China (No. 11920101005, No. 11888101, No. 12047503, No. 12322405, No. 12104450), and the New Cornerstone Investigator Program. C.P. acknowledges financial support from Shanghai Key Laboratory Novel Extreme Condition Materials, China (Grant No. 22dz2260800), Shanghai Science and Technology Committee, China (Grant No. 22JC1410300) and National Science Foundation of China Grant No. W2431011.

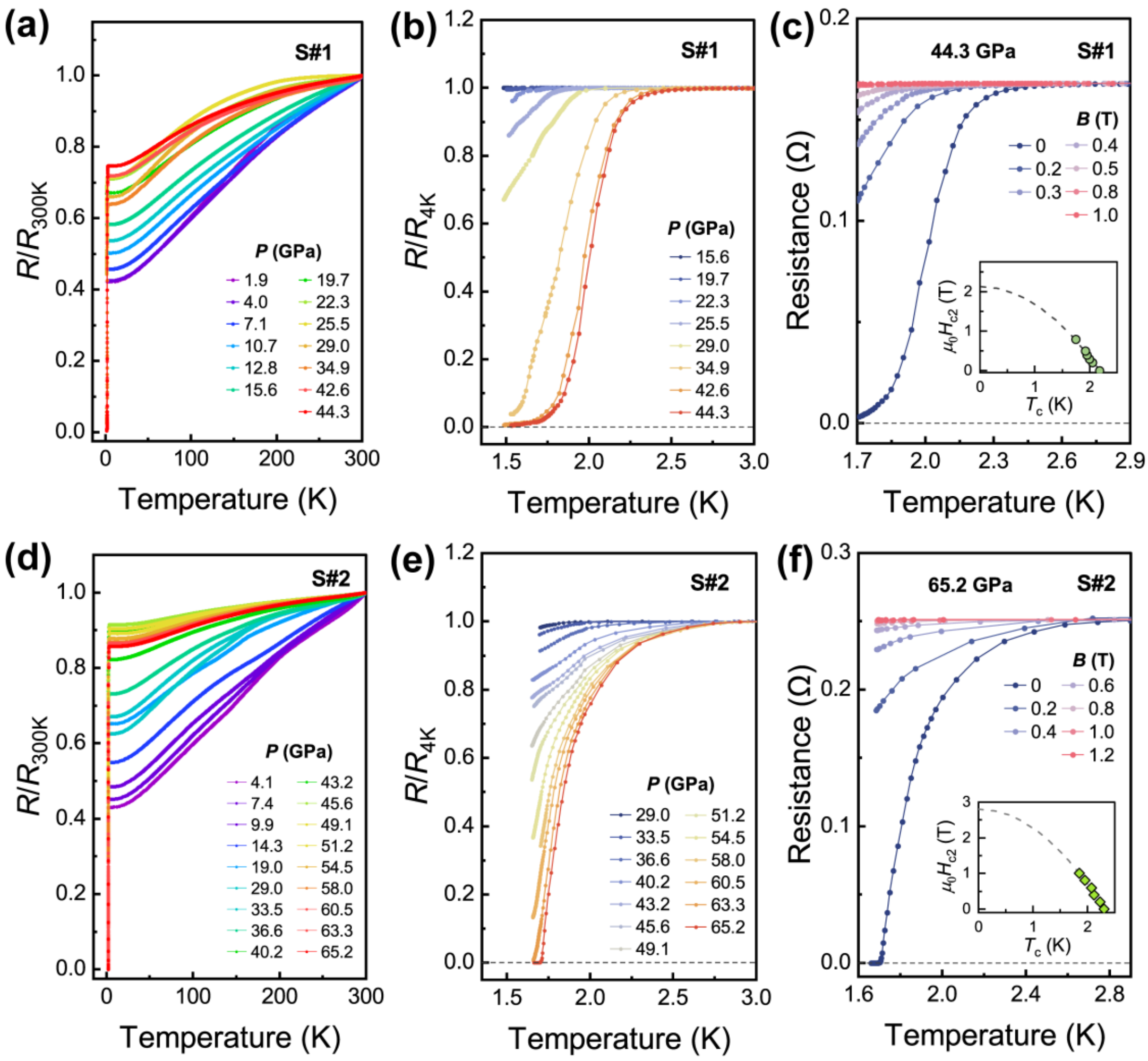

**Fig. 1. Pressure-induced superconductivity in $TaRhTe_4$.** (a) and (d) Temperature dependence of the electrical resistance normalized by its value at 300 K, $R/R_{300\,K}$, for samples S#1 and S#2, respectively, measured at different pressures. (b) and (e) Enlarged low-temperature resistance curves normalized by the resistance at 4 K, $R/R_{4K}$, showing the pressure evolution of the superconducting transition in S#1 and S#2. A zero-resistance state is achieved in S#2 at 63.3 GPa and persists to 65.2 GPa. (c) and (f) Magnetic-field dependence of the superconducting transitions measured at 44.3 GPa for S#1 and 65.2 GPa for S#2, respectively. The transitions shift systematically to lower temperatures with increasing magnetic field. Insets show the temperature dependence of the upper critical field, $\mu_0 H_{c2}(T)$; dashed curves are fits to the Ginzburg-Landau model.

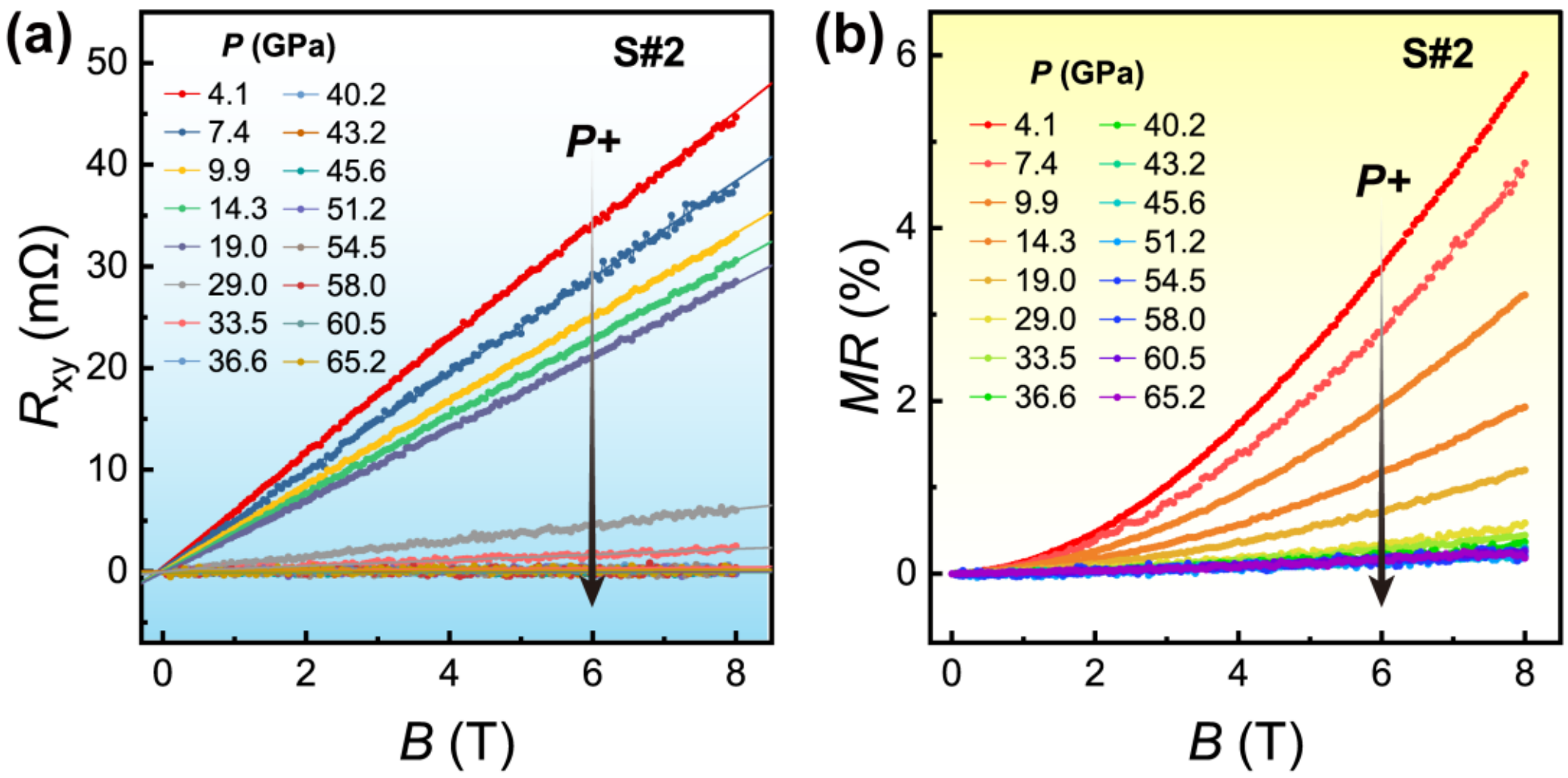

**Fig. 2. Pressure evolution of the Hall resistance and magnetoresistance in $TaRhTe_4$.** (a) Magnetic-field dependence of the Hall resistance $R_{xy}$ of S#2 measured at 10 K under different pressures up to 65.2 GPa. The Hall response is strongly suppressed with increasing pressure and approaches zero in the high-pressure regime. (b) Magnetic-field dependence of the magnetoresistance, defined as $\mathrm{MR}=[R(B)-R(0)]/R(0)\times100\%$, measured under the same conditions. The positive MR decreases rapidly upon compression and remains small at high pressures. The arrows indicate the direction of increasing pressure.

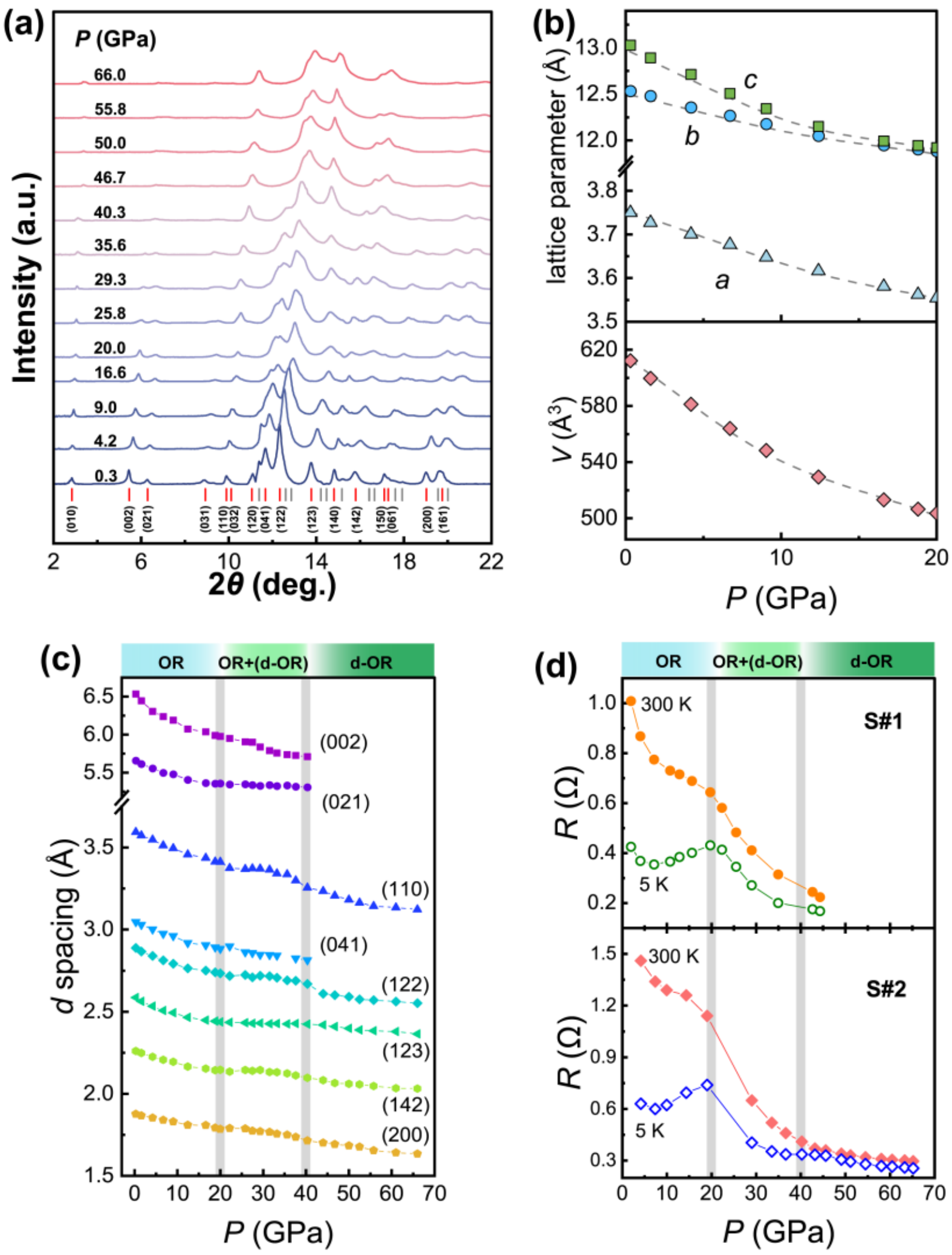

**Fig. 3. Pressure evolution of the crystal structure and electrical resistance of $TaRhTe_4$.** (a) Synchrotron X-ray diffraction patterns collected at different pressures up to 66.0 GPa. Tick marks indicate the calculated reflections of the ambient-pressure orthorhombic structure. (b) Pressure dependence of the lattice parameters *a*, *b*, and *c*, together with the unit-cell volume *V*, in the low-pressure orthorhombic phase. The dashed lines are guides to the eye. (c) Pressure dependence of the *d*-spacings of selected reflections. The pressure range is divided into three regimes: orthorhombic (OR) phase, OR-to-distorted-orthorhombic (d-OR) crossover, and d-OR. (d) Pressure dependence of the electrical resistance at 300 K and 5 K for samples S#1 and S#2. The gray shaded regions in (c) and (d) mark the structural crossover intervals near $P_{c1}$ and $P_{c2}$, respectively.

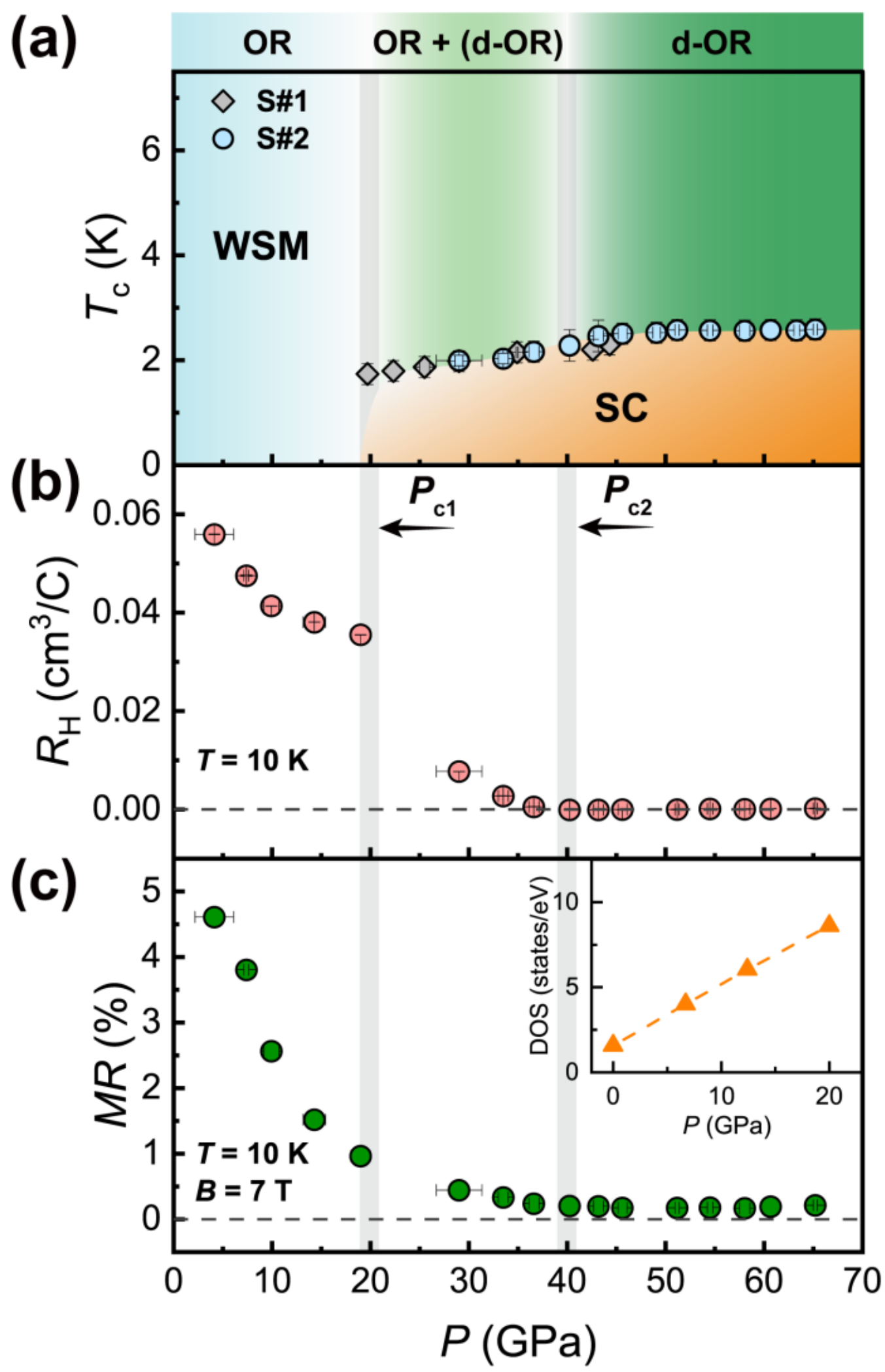

**Fig. 4. Summary of the pressure-induced evolution of superconductivity, crystal structure, and normal-state transport in $TaRhTe_4$.** (a) Pressure-Temperature phase diagram combined with the structural information obtained from high-pressure X-ray diffraction measurements. Diamonds and circles represent the superconducting transition temperatures of S#1 and S#2, respectively. The left denotes the region with orthorhombic (OR) phase, the middle refers to the region of crossover from the OR phase to the distorted-orthorhombic (d-OR) phase, and the left represents the region of d-OR phase.WSM and SC denote the Weyl-semimetal and superconducting states, respectively. (b) Pressure dependence of the Hall coefficient $R_H$, extracted from the linear fits to $R_{xy}(B)$ measured at 10 K. (c) Pressure dependence of the magnetoresistance measured at 10 K and 7 T, where MR=$[R(7\ \text{T})-R(0)]/R(0)\times100\%$. The gray shaded

regions indicate the critical pressures $P_{c1}$ and $P_{c2}$ associated with the structural evolution. The inset shows the calculated total DOS at the Fermi level as a function of pressure.

## References

[1] A. A. Burkov and L. Balents, Weyl Semimetal in a Topological Insulator Multilayer, Phys. Rev. Lett. **107**, 127205 (2011).

[2] N. P. Armitage, E. J. Mele, and A. Vishwanath, Weyl and Dirac semimetals in three-dimensional solids, Rev. Mod. Phys. **90**, 015001 (2018).

[3] A. A. Soluyanov, D. Gresch, Z. Wang, Q. Wu, M. Troyer, X. Dai, and B. A. Bernevig, Type-II Weyl semimetals, Nature **527**, 495 (2015).

[4] X. Wan, A. M. Turner, A. Vishwanath, and S. Y. Savrasov, Topological semimetal and Fermi-arc surface states in the electronic structure of pyrochlore iridates, Phys. Rev. B **83**, 205101 (2011).

[5] D. T. Son and B. Z. Spivak, Chiral anomaly and classical negative magnetoresistance of Weyl metals, Phys. Rev. B **88**, 104412 (2013).

[6] G. Shipunov et al., Layered van der Waals Topological Metals of $TaTMTe_4$ (TM = Ir, Rh, Ru) Family, J. Phys. Chem. Lett. **12**, 6730 (2021).

[7] X. Zhang, N. Mao, O. Janson, J. Van Den Brink, and R. Ray, Layer dependent topological phases and transitions in TaRhTe4 : From monolayer and bilayer to bulk, Phys. Rev. Mater. **8**, 094201 (2024).

[8] H. Rankin, T. J. Slade, B. Schrunk, Y. Kushnirenko, A. Eaton, K. U. R. R. S. Rathnayaka, M. Doyle, L.-L. Wang, P. C. Canfield, and A. Kaminski, *Observation of Flat Bands in Type-II Weyl Semimetal TaRhTe4*, arXiv:2607.01186.

[9] M. Behnami et al., Signature of chiral anomaly in the Weyl semimetal TaRhTe4, Phys. Rev. B **112**, 045101 (2025).

[10] J. Liu, H. Wang, C. Fang, L. Fu, and X. Qian, van der Waals Stacking-Induced Topological Phase Transition in Layered Ternary Transition Metal Chalcogenides,

Nano Lett. **17**, 467 (2017).

[11] Y. Hasuo, T. Urata, T. Hatano, M. Araidai, and H. Ikuta, Intercalation-induced interlayer decoupling in HfTiTe4 and TaRhTe4, Phys. Rev. Mater. **10**, 014201 (2026).

[12] D. Kang et al., Superconductivity emerging from a suppressed large magnetoresistant state in tungsten ditelluride, Nat. Commun. **6**, 7804 (2015).

[13] X.-C. Pan et al., Pressure-driven dome-shaped superconductivity and electronic structural evolution in tungsten ditelluride, Nat. Commun. **6**, 7805 (2015).

[14] P. Lu et al., Origin of superconductivity in the Weyl semimetal WTe2 under pressure, Phys. Rev. B **94**, 224512 (2016).

[15] Y. Qi et al., Superconductivity in Weyl semimetal candidate MoTe2, Nat. Commun. **7**, 11038 (2016).

[16] Y. Li et al., Concurrence of superconductivity and structure transition in Weyl semimetal TaP under pressure, Npj Quantum Mater. **2**, 66 (2017).

[17] J. Zhang, Q. Li, C. Yang, and W. Rao, Evolution of atomic and electronic structures of TaP under high pressure, Comput. Mater. Sci. **142**, 320 (2018).

[18] S. Cai, E. Emmanouilidou, J. Guo, X. Li, Y. Li, K. Yang, A. Li, Q. Wu, N. Ni, and L. Sun, Observation of superconductivity in the pressurized Weyl-semimetal candidate TaIrTe4, Phys. Rev. B **99**, 020503 (2019).

[19] S. Long et al., Observation of nearly identical superconducting transition temperatures in the pressurized Weyl semimetals MIrTe4 ( M = Nb and Ta), Phys. Rev. B **104**, 144503 (2021).

[20] Q.-G. Mu, F.-R. Fan, H. Borrmann, W. Schnelle, Y. Sun, C. Felser, and S. Medvedev, Pressure-induced superconductivity and modification of Fermi surface in type-II Weyl semimetal NbIrTe4, Npj Quantum Mater. **6**, 55 (2021).

[21] M. Jin et al., Discovery of Dome-Shaped Superconducting Phase and Anisotropic Transport in a van der Waals Layered Candidate $NbIrTe_4$ under Pressure, Adv. Sci. **8**, 2103250 (2021).

[22] See Supplemental Material at *** for additional experimental details, data analysis,

and supporting figures.

[23] M. Behnami, Transport Investigation of Topological Materials, Ph.D. thesis, Technische Universität Dresden, 2025.